\documentclass[twocolumn,amsmath,showpacs,amsfonts,aps,prc,floatfix]{revtex4}
\usepackage{graphicx}
\usepackage{bm}
\newcommand{\etas}{\left (\frac{\eta}{s}\right )_{qgp}}
\newcommand{\etah}{\left (\frac{\eta}{s}\right )_{had}}
\begin{document}

\title{Temperature dependence of QGP viscosity over entropy ratio from hydrodynamical analysis of ALICE data in $\sqrt{s_{NN}}$=2.76 TeV Pb+Pb collisions}

\author{A. K. Chaudhuri}
\email[E-mail:]{akc@veccal.ernet.in}
\affiliation{Variable Energy Cyclotron Centre, 1/AF, Bidhan Nagar, 
Kolkata 700~064, India}

\begin{abstract}

Within Israel-Stewart's theory of dissipative hydrodynamics, we have analyzed    ALICE data for the centrality dependence of charged particles multiplicity, elliptic flow and  $p_T$ spectra      in $\sqrt{s}_{NN}$=2.76 TeV Pb+Pb collisions and obtained the temperature dependence of the QGP viscosity over the entropy ratio ($\etas$). 
If temperature dependence of $\etas$ is parameterized as $\etas=\alpha \frac{T-T_c}{T_c}+\frac{1}{4\pi}$,
experimental data  favor $\alpha$ in the range 0-0.2.  $\alpha \geq 0.4$ is not favored  by the data. Experimental data in $\sqrt{s}_{NN}$=200 GeV Au+Au collisions however, prefer $\alpha$=0.4.
 \end{abstract}


\date{\today}  

\maketitle

 \section{Introduction} 
 
In recent years there is considerable interest in QGP viscosity. 
Lattice simulations of QCD predicted that the strongly interacting nuclear matter   undergoes a confinement-deconfinement cross-over transition  \cite{Karsch:2008fe,Cheng:2009zi,Aoki:2006we,Fodor:2010zz} above a critical temperature $T_c\approx 170$ MeV. $\sqrt{s_{NN}}$=200 GeV Au+Au collisions   at Relativistic Heavy Ion Collider (RHIC)  \cite{BRAHMSwhitepaper,PHOBOSwhitepaper,PHENIXwhitepaper,STARwhitepaper} and    $\sqrt{s_{NN}}$=2.76 TeV Pb+Pb collisions at Large Hadron Collider (LHC) \cite{Aamodt:2010pb,Collaboration:2010cz,Aamodt:2010jd,Aamodt:2010pa}, produced convincing evidences that in
RHIC and LHC collisions a deconfined medium or Quark-Gluon Plasma (QGP)  is produced. It is then important to characterize QGP in terms of its transport coefficients, e.g. shear viscosity, bulk viscosity and conductivity. Present paradigm is that shear viscosity over entropy ratio of any matter has a minimum, possibly with a cusp, around the critical temperature  $T=T_c$ \cite{Csernai:2006zz}. 
 String theory based models (AdS/CFT) give a lower bound on viscosity of any matter $\eta/s \geq 1/4\pi$ \cite{Policastro:2001yc,Kovtun:2003wp}. 
 QGP viscosity over entropy ratio ($\etas$) is largely uncertain.
 In a perturbative QCD, Arnold et al  \cite{Arnold:2000dr} estimated $\etas\sim$ 1. 
In a SU(3) gauge theory, Meyer \cite{Meyer:2007ic} gave the upper bound $\etas<$ 1, and his best estimate is $\eta/s$=0.134(33) at $T=1.165T_c$. 
At RHIC region, Nakamura and Sakai \cite{Nakamura:2005yf}
estimated the viscosity of a hot gluon gas  as $\eta/s$=0.1-0.4. In Fig.\ref{F1}, Nakamura and Sakai's \cite{Nakamura:2005yf} estimate of $\etas$  is shown. 
The predictions have large uncertainty and it is difficult to conclude about the temperature dependence
of QGP viscosity. 

In   hydrodynamic models, several authors have estimated viscosity over the entropy ratio of the fluid produced in Au+Au collisions at RHIC and in Pb+Pb collisions at LHC \cite{Luzum:2008cw,Song:2008hj,Drescher:2007cd,Lacey:2006bc,arXiv:0909.0391,arXiv:0909.0376,arXiv:0910.0979,arXiv:1103.2870,Song:2011qa,Shen:2011kn}. However, most of the studies ignore the temperature dependence of viscosity over entropy ratio. 
Recently, in \cite{Niemi:2011ix,arXiv:1002.2394}, temperature dependent viscosity over entropy ratio was considered. Parameterized lattice QCD results \cite{Nakamura:2005yf} for $\etas$ was used. 
It was found that the elliptic flow in
$\sqrt{s}_{NN}$ = 200 GeV Au+Au collisions at RHIC is dominated
by the viscosity in the hadronic phase. It is largely insensitive to QGP viscosity.
QGP viscosity dominates only at  LHC energy $\sqrt{s}_{NN}$ = 2.76 TeV. Temperature dependence of QGP viscosity
was studied in \cite{Song:2011qa,Shen:2011kn}. It was concluded that the temperature dependence of QGP viscosity over entropy ratio cannot be constrained by fitting   spectra and elliptic flow data alone. 

Compared to Au+Au collisions at RHIC, in Pb+Pb collisions at LHC, fluid is produced at higher temperature. QGP phase is much longer at LHC than at RHIC
energy collisions.    Temperature dependence of QGP viscosity over entropy ratio is then better determined from ALICE data.  
In the present paper, we analyze
 the ALICE data  on the centrality dependence of charged particles multiplicity, integrated and differential elliptic flow \cite{Aamodt:2010pb,Collaboration:2010cz,Aamodt:2010jd,Aamodt:2010pa} in $\sqrt{s}_{NN}$=2.76 TeV Pb+Pb collisions and 
  determine  the temperature dependence of QGP viscosity over entropy ratio. 
Fluid is produced progressively at lower temperature as the collisions become more and more peripheral. Centrality dependence of charged particles multiplicity or elliptic flow will have the imprint of initial temperature and consequently will be sensitive to the temperature dependence of viscosity. Our strategy is simple,
we examine whether a given  temperature dependent $\etas$ 
is consistent with the experimental data. 

\section{Hydrodynamic equations, equation of state, initial conditions and temperature dependence of $\etas$}

We assume that in $\sqrt{s}_{NN}$=2.76 TeV, Pb+Pb collisions   a baryon free fluid is formed. The space-time evolution of the fluid is obtained by solving Israel-Stewart's 2nd order hydrodynamic equations,

 \begin{eqnarray}  
\partial_\mu T^{\mu\nu} & = & 0,  \label{eq3} \\
D\pi^{\mu\nu} & = & -\frac{1}{\tau_\pi} (\pi^{\mu\nu}-2\eta \nabla^{<\mu} u^{\nu>}) \nonumber \\
&-&[u^\mu\pi^{\nu\lambda}+u^\nu\pi^{\nu\lambda}]Du_\lambda. \label{eq4}
\end{eqnarray}

Eq.\ref{eq3} is the conservation equation for the energy-momentum tensor, $T^{\mu\nu}=(\varepsilon+p)u^\mu u^\nu - pg^{\mu\nu}+\pi^{\mu\nu}$, 
$\varepsilon$, $p$ and $u$ being the energy density, pressure and fluid velocity respectively.   Eq.\ref{eq4} is the relaxation equation for the shear stress tensor $\pi^{\mu\nu}$.   
In Eq.\ref{eq4}, $D=u^\mu \partial_\mu$ is the convective time derivative, $\nabla^{<\mu} u^{\nu>}= \frac{1}{2}(\nabla^\mu u^\nu + \nabla^\nu u^\mu)-\frac{1}{3}  
(\partial . u) (g^{\mu\nu}-u^\mu u^\nu)$ is a symmetric traceless tensor. $\eta$ is the shear viscosity and $\tau_\pi$ is the relaxation time.  It may be mentioned that in a conformally symmetric fluid relaxation equation can contain additional terms  \cite{Song:2008si}.    
Assuming boost-invariance, Eqs.\ref{eq3} and \ref{eq4}  are solved in $(\tau=\sqrt{t^2-z^2},x,y,\eta_s=\frac{1}{2}\ln\frac{t+z}{t-z})$ coordinates, with the code 
  "`AZHYDRO-KOLKATA"', developed at the Cyclotron Centre, Kolkata.
 Details of the code can be found in \cite{Chaudhuri:2008sj}. 
 
Hydrodynamic equations  are closed with an equation of state (EoS) $p=p(\varepsilon)$.
In the present study, we use an equation of state  where the Wuppertal-Budapest \cite{Aoki:2006we}
lattice simulations for the deconfined phase is smoothly joined at $T=T_c=174$ MeV, with hadronic resonance gas EoS comprising all the resonances below mass $m_{res}$=2.5 GeV. Details of the EoS can be found in \cite{arXiv:1103.2870}. 
 
Solution of partial differential equations (Eqs.\ref{eq3},\ref{eq4}) requires
 energy momentum tensor $T^{\mu\nu}$ and shear stress tensor $\pi^{\mu\nu}$ at the transverse plane at the initial time $\tau_i$. We assume that   in impact parameter {\bf b} collision, at the initial time $\tau_i$=0.6 fm, energy density is proportional to the transverse profile of the average number of binary collisions $N_{coll}({\bf b},x,y)$, calculated in a Glauber model,
 
\begin{equation} \label{eq5}
\varepsilon({\bf b},x,y)=\varepsilon_i N_{coll}({\bf b},x,y)],
\end{equation}

$\varepsilon_i$ is the central energy density in a zero impact parameter collisions. We also assume zero initial fluid velocity, $v_x(x,y)=v_y(x,y)=0$. In dissipative hydrodynamics, apart from the energy density and
fluid velocity, shear stress tensor has to be initialised, We initialise the shear stress tensors  to the boost-invariant values, $\pi^{xx}=\pi^{yy}=2\eta/3\tau_i$, $\pi^{xy}$=0, $\eta$ being the shear viscosity coefficient. We have to specify $\eta$ as well as the relaxation time $\tau_\pi$. For the  
relaxation time we use the   Boltzmann estimate $\tau_\pi=3\eta/2p$. Shear viscosity coefficient is parameterized as follows:
we assume that QGP viscosity over entropy ratio $\etas$ has a minimum at $T=T_c$ and to be consistent with AdS/CFT conjecture we take $1/4\pi$ as  the minimum value. We approximate the temperature dependence of viscosity over entropy ratio of the QGP fluid   by a linear relation, 

  \begin{figure}[t]
\center
 \resizebox{0.30\textwidth}{!}{%
  \includegraphics{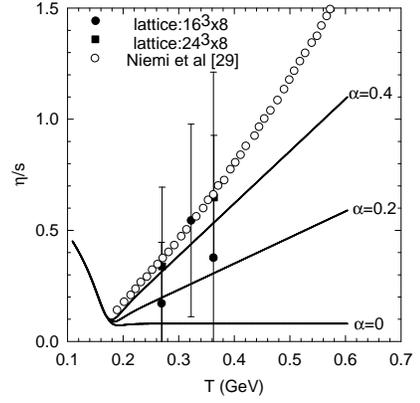}
}
\caption{Three forms of temperature dependence of viscosity over the entropy ratio ($\etas$), used in the present simulations are shown. The black circles and squares are lattice QCD predictions \cite{Nakamura:2005yf} for the QGP viscosity over entropy ratio. The while circles are approximate temperature dependence of $\etas$ used in \cite{Niemi:2011ix}.}
  \label{F1}
\end{figure}  

\begin{equation} \label{eq1}
\left ( \frac{\eta}{s} \right )_{qgp} (T) =\alpha\frac{T-T_c}{T_c} + \frac{1}{4\pi}, 
\end{equation}

\noindent the parameter $\alpha$ control the temperature dependence.
Recognizing the uncertainty in  the temperature dependence,  we have considered three values for $\alpha$, (i) stiff temperature dependence: $\alpha$=0.4 (approximately consistent with lattice simulations), (ii) moderate temperature dependence: $\alpha$=0.2 and (iii) temperature independence: $\alpha$=0. Below the critical temperature $T_c$, QGP fluid crosses over to hadronic fluid.    For the hadronic fluid viscosity over entropy ratio $(\frac{\eta}{s})_{hrg}$  we use the results for hadronic resonance gas \cite{NoronhaHostler:2008ju}. The two viscosities are smoothly joined at $T=T_c$,

\begin{eqnarray} \label{eq2}
\frac{\eta}{s}(T)&=&\frac{1}{2} \left [1-\tanh\frac{T-T_c}{\Delta T}\right ]
\left (\frac{\eta}{s}\right )_{hrg}(T)\\
&+& \frac{1}{2}\left [1+\tanh\frac{T-T_c}{\Delta T}\right ]
\left (\frac{\eta}{s}\right )_{qgp}(T) \nonumber
\end{eqnarray}

Particle production is not very sensitive to the smoothening parameter $\Delta T$. In the following we use $\Delta T$=0.02 GeV. In Fig.\ref{F1}, 
temperature dependence of $\frac{\eta}{s}$ for the three values of $\alpha$ is shown.

We use the 
Cooper-Frye prescription to obtain the invariant distributions of hadrons, from the freeze-out surface at $T_F$=130 MeV.
In Cooper-Frye, invariant distribution is obtained as,

\begin{equation}
E\frac{dN}{d^3p}=\int_\Sigma d\Sigma_\mu p^\mu f(x,p)
\end{equation}

\noindent where $\Sigma_\mu$ is the freeze-out hypersurface and f(x,p) is the distribution function. In viscous hydrodynamics, $f(x,p)$ is a non-equilibrium distribution function.   Assuming that the fluid is not far from equilibrium, we approximate $f(x,p)$ as ,

\begin{equation}f(x,p)=f_{eq}(x,p) [1+\phi(x,p)],\end{equation}

\noindent where $f_{eq}(x,p)$ is the equilibrium distribution function. The deviation function $\phi(x,p)$ is approximated as, 

\begin{equation}
\phi(x,p)=\frac{1} {2(\varepsilon+P)T^2} p_\mu p_\nu \pi^{\mu\nu} << 1,
\end{equation}

\noindent where, $\varepsilon$, $P$ and $T$ are (local) energy density, pressure and temperature respectively.

 \begin{figure}[t]
\center
 \resizebox{0.30\textwidth}{!}{%
  \includegraphics{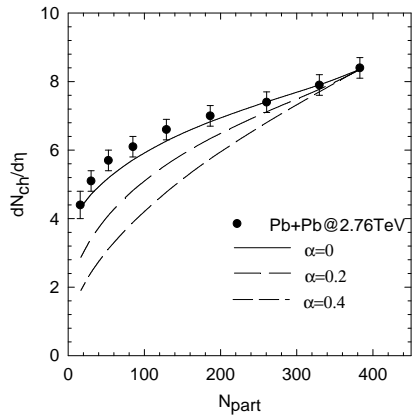}
}
\caption{The black circles are ALICE data for centrality dependence of charged particles multiplicity per participant pair in   $\sqrt{s}_{NN}$= 2.76 TeV Pb+Pb collisions \cite{Collaboration:2010cz}. The solid, long dashed, short dashed   lines are hydrodynamic simulations with the three forms of $\etas$ (see Eq.\ref{eq1}).
}
  \label{F2}
\end{figure}

 We may mention here that there is some overlap between the main idea in the present paper and that of Ref.\cite{Niemi:2011ix}. We emphasize the differences here. In \cite{Niemi:2011ix}, effect of temperature dependent viscosity over entropy ratio, on elliptic flow and transverse momentum spectra was investigated.
A fixed temperature dependent $\etas$, from lattice QCD calculations, was used in the simulations. In the present simulations, on the other hand, we try to determine the temperature dependence of $\etas$. Three different temperature dependence of $\etas$ is tested against the experimental data. As shown in Fig.\ref{F1},  
temperature dependence of $\etas$ in \cite{Niemi:2011ix} is steeper than the steepest QGP viscosity over 
entropy ratio used presently. There are some minor differences in details of the simulation also, e.g. in
equations of state, initialisation of shear stress tensor, decoupling temperature etc.

 
    \begin{figure}[t]
\center
 \resizebox{0.30\textwidth}{!}{%
  \includegraphics{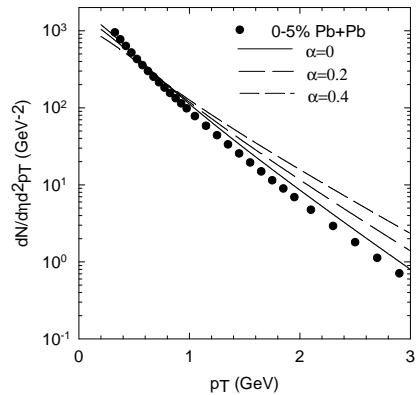}
}
\caption{  Black   circles are charged particle's spectra in 0-5\%   Pb+Pb collisions   \cite{Aamodt:2010jd}. The solid, long dashed, and short dashed lines are hydrodynamic model simulations for  charged particle's spectra with the three forms of $\etas$ (see Eq.\ref{eq1}). 
}\label{F3}
\end{figure}   
\section{Results}
 
\subsection{Pb+Pb collisions at LHC}

\begin{table}[h]
\caption{\label{table1} Central energy density ($\varepsilon_i$) and temperature ($T_i$) required to reproduce experimental charged particle's multiplicity in 0-5\% Pb+Pb collisions at $\sqrt{s}_{NN}$=2.76 TeV at LHC. We have also listed $\varepsilon_i$ and $T_i$ (the bracketed quantities) required to reproduce the experimental charged particle's multiplicity in 0-5\% Au+Au collisions at $\sqrt{s}_{NN}$=200 GeV at RHIC.   Viscosity over the entropy ratio of the central fluid is also noted. }
\begin{ruledtabular} 
  \begin{tabular}{|c|c|c|c|}\hline
  & $\varepsilon_i $ & $T_i $ &$\left (\frac{\eta}{s}\right )_{qgp}$\\ 
  & $(GeV/fm^3)$ & $(MeV)$ & \\ \hline
$\alpha=0$  & $135\pm7$ & $541\pm 7$ & 0.08\\
            & ($32\pm 2$) & ($380\pm 6$) & (0.08)  \\ \hline
$\alpha=0.2$  & $78\pm 4$  & $473\pm 6$ & 0.43 \\ 
              & ($22\pm 2.0$)  & ($345\pm 6$) & (0.20)  \\ \hline
$\alpha=0.4$  & $49\pm 2$  &$422\pm 4$ &0.96\\
              & ($16\pm 1$) & ($319\pm 5$) & (0.34)\\
\end{tabular}\end{ruledtabular}  
\end{table}

The central energy density $\varepsilon_i$ of    the fluid, at the initial time $\tau_i$=0.6 fm, is the only free parameter in the model.
We fix $\varepsilon_i$ to reproduce experimental charged particle's multiplicity $\frac{dN_{ch}}{d\eta}=1601\pm 60$ in 0-5\% Pb+Pb collision \cite{Collaboration:2010cz}. For the three forms of QGP viscosity, $\alpha$=0.4, 0.2 and 0, we have simulated 0-5\% Pb+Pb collisions and computed negative pion multiplicity. Resonance production is included. Noting that pion's constitute $\approx$80\% of the total charged particles, $\pi^-$ multiplicity is multiplied by the factor $2\times 1.20$ to compare with experimental charged particle multiplicity.   In table.\ref{table1}, central energy density and temperature required to reproduce experimental   multiplicity in 0-5\% collisions are listed. Uncertainty in $\varepsilon_i$ or $T_i$ reflect the   uncertainty in ALICE measurements.  As expected, central energy density or temperature of the fluid is reduced with increasing value of $\alpha$. Charged particles multiplicity is a measure of the final state entropy. In viscous fluid, during evolution, entropy is generated. Entropy generation increases with viscosity and more viscous fluid require less initial energy density to reach the fixed final state entropy.  
In table.\ref{table1}, we have also listed the viscosity over the entropy ratio of the central fluid. One note that central fluid viscosity is very large of  if $\etas$ has stiff temperature dependence.

However,  stiff temperature dependence of $\etas$   is not consistent with the experimental centrality dependence of charged particles multiplicity.  In Fig.\ref{F2} simulated charged particles multiplicity per participant pairs ($\frac{1}{.5N_{part}}\frac{dN_{ch}}{d\eta}$) for the three values of $\alpha$,   are compared with  the ALICE measurements \cite{Collaboration:2010cz}.  
ALICE data for the centrality dependence of charged particles multiplicity is not reproduced for $\alpha$=0.4. In mid-central and peripheral collisions, simulation produces fewer particles per participant pairs  than in experiment.  For moderate temperature dependence, $\alpha$=0.2, agreement with experimental data in mid-central collisions improves but data in peripheral collisions are still under predicted. Of the three forms, data are best explained for $\alpha$=0, when $\etas$   is temperature independent. ALICE data on the centrality dependence of charged particles multiplicity do not demand temperature dependent QGP viscosity over entropy ratio. Rather, it is consistent with the AdS/CFT minimal value $\etas=\frac{1}{4\pi}$.


    \begin{figure}[t]
\center
 \resizebox{0.30\textwidth}{!}{%
  \includegraphics{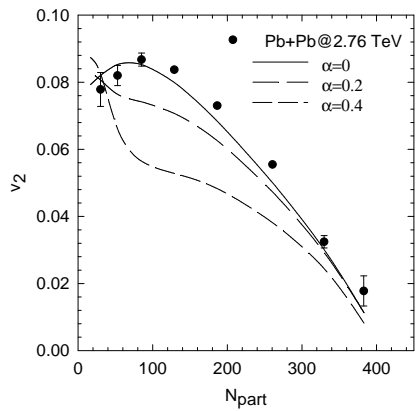}
}
\caption{   ALICE measurement \cite{Aamodt:2010pa} for charged particle's elliptic flow in   4-particle cumulant method are compared with the simulated flows   for three forms of $\etas$ (see Eq.\ref{eq1}). 
}
  \label{F4}
\end{figure}

Temperature independent $\etas$ also best explains he ALICE measurements for the charged particle $p_T$ spectra in 0-5\% Pb+Pb collisions \cite{Aamodt:2010jd}. In Fig.\ref{F3}, the simulated charged particles spectra,  for the three values of $\alpha$, $\alpha$=0, 0.2 and 0.4, are plotted against the ALICE data. Note that all the three curves reproduce the charged particles multiplicity (see Fig.\ref{F2}). Spectra become flatter as $\alpha$ increases. ALICE data are best explained when $\etas$ is temperature independent ($\alpha$=0). Even then, we do note that the model over predict the spectra at $p_T>$ 1 GeV. 
The reason is understood. We have approximated charged particles spectra by the pion spectra. However, experimentally, kaon and proton spectra are flatter than the pion spectra. When constraint to reproduce the total multiplicity, the approximation will overestimate the central fluid temperature, increasing the high $p_T$ yield.

 Integrated elliptic flow ($v_2$) is an important observable in relativistic energy collisions. It is a measure of collectivity in the medium. In Fig.\ref{F4},
 ALICE measurements \cite{Aamodt:2010pa} for $v_2$ is shown. As the collisions become more and more peripheral, $v_2$ or collectivity increases, till very peripheral collisions where we observe a dip in collectivity. In Fig.\ref{F4}, solid, long dashed and short dashed lines are  simulated (integrated) elliptic flow for $\alpha$=0, 0.2 and 0.4 respectively. For $\alpha$=0.4, flow is largely under predicted. As it was for charged particles multiplicity, centrality dependence of elliptic flow also donor require stiff temperature dependent $\etas$. Though the agreement with data is better for $\alpha$=0.2, the model predictions under predict the data.  Data are best described when $\alpha$=0, i.e. $\etas$ is temperature independent.  

   \begin{figure}[t]
\center
 \resizebox{0.35\textwidth}{!}{%
  \includegraphics{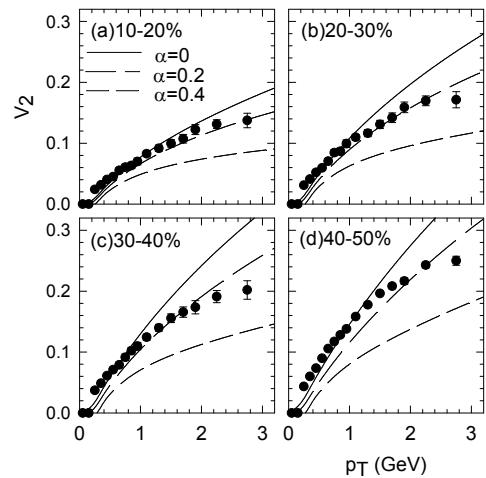}
}
\caption{ The black circles  are  the ALICE measurement \cite{Aamodt:2010pa} for charged particle's elliptic flow in 0-10\%, 10-20\%, 20-30\% and 30-40\% Pb+Pb collisions. The solid, long dashed and short dashed lines are hydrodynamical model simulations with the three forms of $\etas$ (see Eq.\ref{eq1}). 
}
  \label{F5}
\end{figure}  
 
A different conclusion is reached when we consider centrality dependence of differential elliptic flow ($v_2(p_T)$).  In Fig.\ref{F5}, ALICE measurements for the charged particle's  differential elliptic flow in 10-20\%, 20-30\%, 30-40\% and 40-50\% Pb+Pb collisions are compared with the simulated differential flows. 
ALICE data on differential flow also donor demand   stiff temperature dependent $\etas$. For $\alpha$=0.4 simulated flows largely  under predict experiment. Data are best explained when $\alpha$=0.2, i.e. for moderate temperature dependence of $\etas$. Temperature independent $\etas$ give comparatively poor description.

 \begin{figure}[t]
\center
 \resizebox{0.35\textwidth}{!}{%
  \includegraphics{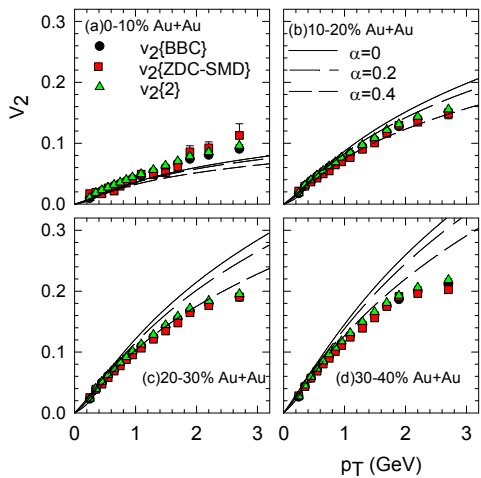}
}
\caption{(color online)
In four panels, PHENIX data \cite{Afanasiev:2009wq}  for the charged particle's (differential) elliptic flow   in 0-10\%, 10-20\%, 20-30\% and 30-40\% Au+Au collisions at     $\sqrt{s}_{NN}$= 200 GeV Au+Au collisions are shown. The solid, long dashed, short dashed   lines in each panels are the hydrodynamic simulations for the $p_T$ spectra with the three forms of $\etas$ (see Eq.\ref{eq1}).
}
  \label{F6}
\end{figure}

\subsection{Au+Au collisions at RHIC} 

Present analysis of ALICE data in $\sqrt{s}_{NN}$=2.76 TeV Pb+Pb collisions clearly indicate that the data disfavor strong temperature dependent $\etas$. A similar analysis can be performed for the experimental data in RHIC energy collisions. It is  interesting to see whether or not QGP produced at RHIC also disfavors strong temperature dependent $\etas$. As before, in $\sqrt{s}_{NN}$=200 GeV Au+Au collisions also, we fix the central energy density $\varepsilon_i$ of the fluid to reproduce the charged particle's multiplicity $\frac{dN_{ch}}{d\eta}=687\pm 37$, in 0-5\% Au+Au collisions \cite{Adler:2004zn}. They are listed in table.\ref{table1}.
Compared to LHC energy, in RHIC energy collisions, fluid is produced at a lower temperature. For example, for temperature independent QGP viscosity ($\alpha$=0),  the central fluid temperature is $T_i\approx$380 MeV, to be contrasted with $T_i\approx$540 MeV in LHC energy collisions. For stiff temperature dependent QGP viscosity ($\alpha$=0.4) the central fluid temperature is $T_i\approx$320 MeV, approximately 100 MeV lower than the fluid temperature at LHC energy collisions. 
QGP produced in RHIC energy collisions then explore more restricted
temperature range than at LHC energy. Consequently, effect of viscosity will be comparatively less at RHIC than at LHC.

 In Fig.\ref{F6}, PHENIX measurements \cite{Afanasiev:2009wq} for the charged particles (differential) elliptic flow in 0-10\%, 10-20\%, 20-30\% and 30-40\% Au+Au collisions are compared with the simulated flows. The solid, dashed and long dashed lines are present model simulations with the  temperature dependence controlling parameter $\alpha$=0, 0.2 and 0.4 respectively.  As expected, elliptic flow is reduced as the temperature dependence of $\etas$ become stiffer. 
 However, reduction is much less than in Pb+Pb collisions at LHC energy. For example, in Pb+Pb collisions at LHC, in 10-20\% collision, at $p_T$=2 GeV, simulated flow is reduced by $\sim$ 20\% when   $\alpha$ is increased from 0 to 0.2 and by $\sim$ 50\% when $\alpha$ is increased from 0 to 0.4. 
  At RHIC energy however, the reduction is much less, $\sim$6\% and $\sim$20\% as $\alpha$ is increased from 0 to 0.2 and 0 to 0.4. Effect of viscosity is less on RHIC data due to reduced fluid temperature.
   Fits obtained to the experimental flow are also interesting.  
In central 0-10\% collisions, for all the three forms of $\etas$, reproduces the experimental data upto $p_T \leq$1.5 GeV. At higher $p_T$ data are under predicted.  In 10-20\% collisions, PHENIX data are marginally over predicted in simulation with temperature independent $\etas$. Simulations with moderate or stiff temperature dependent $\etas$ better describe the data. In 20-30\% or in 30-40\% collisions also, temperature independent $\etas$ over predict the data at large $p_T$. Data are best explained with stiff temperature dependent $\etas$.  
 We have not shown here, but our analysis of charged particles $p_T$ spectra in Au+Au collisions at RHIC also favor stiff temperature dependent $\etas$, rather than temperature independent or moderate temperature dependent $\etas$ . Unlike in Pb+Pb collisions at LHC, in Au+Au collisions at RHIC, experimental   data prefer   strong temperature dependent $\etas$.

 Present analysis of experimental data in RHIC and LHC energy collisions indicate that a unique temperature dependent $\etas$ does not explains the experimental data both at RHIC and LHC energy.  In LHC energy collisions, experimental data prefer temperature independent ($\alpha$=0) or moderate  temperature dependent ($\alpha$=0.2)  $\etas$, but in RHIC energy collisions, data prefer strong temperature dependent ($\alpha$=0.4) $\etas$.
Results of the present  analysis   are similar to that of \cite{Song:2011qa}.
In \cite{Song:2011qa}, a hybrid model VISHNU was used to analyse the experimental data at RHIC and LHC energy collisions. In the hybrid model, below a switching temperature ($T_{switch}$=165 MeV), evolution of the fluid is governed by the  Israel-Stewart's 2nd order hydrodynamic equations. Above the switching temperature, the evolution is obtained using the microscopic hadronic transport model from UrQMD.  In UrQMD,  the kinetic and chemical freeze-out is governed by the Boltzmann equation and the model thus eliminates these two parameters required in pure hydrodynamic approach.  In \cite{Song:2011qa} it was concluded that the temperature dependence of $\etas$ cannot be determined from fitting spectra or elliptic flow data alone. 
The present,  pure hydrodynamic model analysis also indicate that an unique temperature dependent QGP viscosity over entropy ratio do not explains the experimental data both   at RHIC and LHC energy.

\section{Summary and conclusions}

In Israel-Stewart's 2nd order hydrodynamics, we have analysed the ALICE data in $\sqrt{s}_{NN}$=2.76 TeV Pb+Pb collisions to determine the temperature dependence of
QGP viscosity over entropy ratio. Parameterising the temperature dependence of QGP viscosity over entropy ratio as, $\etas=\alpha \frac{T-T_c}{T_c}+\frac{1}{4\pi}$, for three values of $\alpha$, $\alpha$=0, 0.2 and 0.4, we have simulated Pb+Pb collisions and compared with the ALICE data.   ALICE data are best explained for $\alpha$ in the range 0-0.2. Strong temperature   dependence $\alpha \geq 0.4$ is disfavored by the data. In $\sqrt{s}_{NN}$=200 GeV Au+Au collisions, on the other hand, experimental data on elliptic flow or $p_T$ spectra demand stiff temperature dependent $\etas$.

In the present analysis, we have assumed $\etas$ depends linearly on temperature. It is possible that the temperature dependence of $\etas$ is more complex. 
Indeed, one observes that for $\alpha$=0-0.4, differential elliptic flow in Pb+Pb collisions at LHC, or in Au+Au collisions at RHIC, are not properly explained in all the collision centralities. While in $\sqrt{s}_{NN}$=2.76 TeV Pb+Pb collisions $\alpha$=0.4 largely underestimate $v_2(p_T)$, $\alpha$=0-0.2, overestimate $v_2(p_T)$ at   large $p_T$. In $\sqrt{s}_{NN}$=200 GeV Au+Au collisions also,
elliptic flow at large $p_T$ are overestimated in peripheral collisions.  
Linear temperature dependence of $\etas$ may be inadequate for  differential elliptic flow, which a very sensitive observable. Also, for hadronic gas viscosity over entropy ratio $\etah$, we have used the results of \cite{NoronhaHostler:2008ju},  it was not varied.
 Certainly, Ref.\cite{NoronhaHostler:2008ju} does not provide the final word on how large $\eta/s$ is in the hadronic gas phase. Whether or not uncertainties in $\etah$ interferes with extraction of $\etas$ is not investigated. In hydrodynamical models, particle production depends, apart from  viscosity, also on a number of initial conditions, e.g.   initial time, initial energy density/velocity distribution, freeze-out condition etc. We have only varied the central energy density, did not explore all the possible initial conditions.
Correlation between temperature dependence of $\etas$ and various initial conditions, e.g. CGC or fluctuating initial conditions etc. is not studied. Considering all the uncertainties,  
we summarize our analysis as follows: with Glauber model  initialisation,
if the temperature dependence of QGP viscosity over entropy ratio is parameterized as $\etas=\alpha \frac{T-T_c}{T_c}+\frac{1}{4\pi}$,
ALICE data in $\sqrt{s}_{NN}$=2.76 TeV Pb+Pb collisions are best explained for $\alpha$ in the range 0-0.2. Data disfavor strong temperature dependence, $\alpha \geq 0.4$. However, RHIC data on charged particle's elliptic flow or $p_T$ spectra on the other hand prefer $\alpha$=0.4, i.e. strong temperature dependent $\etas$.  
A unique, linear temperature dependent QGP viscosity over entropy ratio 
fails to explain, simultaneously, the experimental elliptic flow data at RHIC and LHC energy.



\end{document}